\documentstyle[preprint,aps]{revtex}
\tighten

\newcommand{\mi}{\text{i}}
\newcommand{\md}{\text{d}}
\newcommand{\e}{\text{e}}
\newcommand{\half}{\case{1}{2}}
\newcommand{\xbj}{x_{\text{B}}}
\newcommand{\tr}{\text{Tr}}
\newcommand{\bkt}{\bbox{k}_T}
\newcommand{\bkat}{\bbox{k}^\prime_T}
\newcommand{\bSt}{\bbox{S}_T}
\newcommand{\bat}{\bbox{a}_T}
\newcommand{\bqt}{\bbox{q}_T}
\newcommand{\bpt}{\bbox{p}_T}
\newcommand{\bap}{\bbox{a}_\perp}
\newcommand{\bPhp}{\bbox{P}_{h\perp}}
\newcommand{\blp}{\bbox{l}_\perp}
\newcommand{\bSp}{\bbox{S}_\perp}
\newcommand{\hh}{\hat{h}}
\newcommand{\bh}{\hat{\bbox{h}}}

\begin{document}

\preprint{\parbox[b]{3.3 cm} {NIKHEF 95-001\\hep-ph/9501202}}

\draft

\title{Probing transverse quark polarization\\
       in deep-inelastic leptoproduction}

\author{R.D. Tangerman and P.J. Mulders\thanks{Also at Physics
Department, Free University, NL-1081~HV Amsterdam, The Netherlands.}}

\address{National Institute for Nuclear Physics and High Energy Physics
(NIKHEF),\\ P.O. Box 41882, NL-1009~DB Amsterdam, The Netherlands }

\date{January 1995}

\maketitle

\begin{abstract}

The azimuthal dependence of hadrons produced in lepton scattering off
a polarized hadron probes the quark transverse-spin distributions.
In the calculation of the asymmetries, transverse momenta of quarks in
the distribution and fragmentation functions must be incorporated.
In addition to the $\sin (\phi + \phi_S)$ asymmetry for transversely
polarized hadrons, known as the Collins effect, we find $\sin 2\phi$
asymmetries for both transversely and longitudinally polarized hadrons.

\end{abstract}


In hard scattering processes one has the possibility of measuring
specific matrix elements of quark and gluon fields. In leading order
they can be readily interpreted as quark and gluon densities including
also the spin (helicity and transverse spin) densities.
Inclusive measurements in deep-inelastic lepton-hadron scattering (DIS)
enable the extraction of the unpolarized distribution $f_1(x)$ and one
of two spin distributions, namely the helicity distribution $g_1(x)$.
The other (transverse) spin distribution $h_1(x)$, cannot be measured
in inclusive DIS because of its chiral structure. It has been suggested
to measure this in Drell-Yan scattering~ \cite{DY} or in
semi-inclusive measurements in lepton-hadron scattering~\cite{SI}.
One of the ways in which semi-inclusive
measurements could be used involves transverse momentum dependence. In
Ref.~\cite{coll93}, Collins shows how the semi-inclusive
deep-inelastic process $e+p^{\uparrow}\rightarrow e+h+X$, where
$p^{\uparrow}$ denotes a transversely polarized hadron
(spin vector $\bSt^i$), enables one to probe the quark transverse spin
through a leading asymmetry depending on the azimuthal angle of the
outgoing hadron's momentum and that of the target hadron's spin vector,
the so-called Collins effect.
In Ref.~\cite{tang94a}, however, we found that terms proportional to
$\bkt^i(\bkt\cdot\bSt/M^2)$ in the relevant matrix element~(\ref{eq:h1})
are neither forbidden by the symmetries of QCD, nor suppressed by powers
of $1/Q$.
In this letter we present {\em all\/} leading contributions that enter
when transverse momentum dependence is considered, treating also the
case where the initial hadron is longitudinally polarized.

We consider the process
$\ell + H \ \longrightarrow \ \ell^\prime + h + X$,
where $\ell$ and $\ell^\prime$ are the incoming and scattered leptons
(considered massless) with momenta $l$ and $l^\prime$,
$H$ is the incoming hadron with momentum $P$ (with $P^2=M^2$) and spin
vector $S$ (with $P\cdot S=0$ and $S^2=-1$),
$h$ is the produced hadron with momentum $P_h$ (with $P_h^2=M_h^2$).
The cross section for this process can be written as a product of a
leptonic and a hadronic tensor
\begin{equation}
\frac{{\md}\sigma}{{\md}\xbj \,{\md}y\,{\md}z\, {\md}^2\bPhp}
= \frac{\pi y\alpha^2}{2zQ^4}2M {\cal W}^{\mu\nu}L_{\mu\nu},
\end{equation}
where we consider the situation in which the transverse momentum
$\bPhp^2$ is assumed to be of
${\cal O}(M^2)$. We have used the scalar variables
\begin{equation}
\xbj=\frac{Q^2}{2P\cdot q},\qquad y=\frac{P\cdot q}{P\cdot l},
\qquad z=\frac{P\cdot P_h}{P\cdot q}.
\end{equation}
Here $q = l-l^\prime$ is the momentum of the exchanged virtual photon
(we will limit ourselves to electromagnetic interactions),
which is spacelike ($-q^2$ = $Q^2$).
We assume that $Q^2$ becomes large as compared to the hadronic scale,
say $M^2$, but $\xbj$ and $z$ remain constant, well away from their
endpoints $0$ and $1$.
It is convenient to introduce the vector $\tilde P^\mu \equiv P^\mu
- (P\cdot q/q^2)\,q^\mu$ which is orthogonal to $q$, leading to the
timelike unit vector $\hat{t}^\mu \equiv$  $2\xbj\,\tilde P^\mu/Q$
(with $\hat{t}^2 = 1$). Together with the spacelike unit vector
$\hat{q}^\mu \equiv$ $q^\mu/Q$ (with $\hat{q}^2 = -1$),
we can define tensors in the space orthogonal to $P$ and $q$,
\begin{eqnarray}
&&g_\perp^{\mu \nu} \equiv
g^{\mu \nu} + \hat{q}^\mu \hat{q}^\nu  - \hat{t}^\mu \hat{t}^\nu
= g^{\mu \nu} - \frac{q^\mu q^\nu}{q^2}
- \frac{\tilde P^\mu \tilde P^\nu}{\tilde P^2} ,
\label{tensorg}\\
&&\epsilon_\perp^{\mu \nu} = \epsilon^{\mu \nu \rho \sigma}\hat{t}_\rho
\hat{q}_\sigma = \frac{2\xbj}{Q^2}\,\epsilon^{\mu \nu \rho \sigma}P_\rho
q_\sigma.
\label{tensore}
\end{eqnarray}
Azimuthal angles will be fixed with respect to the lepton scattering
plane. The $x$-axis is derived from the lepton momentum which can be
written as
\begin{equation}
l^\mu = \frac{(2-y)Q}{2y}\,\hat{t}^\mu + \frac{Q}{2}\,\hat{q}^\mu
+ l_\perp^\mu,
\end{equation}
with $\blp^2$ = $-l_\perp^2$ = $Q^2(1-y)/y^2$. We define the
unit vector in the $x$-direction to be
$\hat{x}^\mu\equiv l_\perp^\mu/| \blp|$.
The (unpolarized) lepton tensor
$L^{\mu\nu}=2l^{\mu} l'^{\nu}+2l^{\nu} l'^{\mu}-Q^2 g^{\mu\nu}$ can be
written as
\begin{eqnarray}
L^{\mu\nu}=\frac{4Q^2}{y^2}\biggl[ &&
-\half\left(1-y+\half y^2\right)\, g_\perp^{\mu\nu}
+(1-y)\,\left(\hat{x}^\mu\hat{x}^\nu+\half g_\perp^{\mu\nu}\right)
\nonumber\\ &&+(1-y)^{\frac{1}{2}}\left(1-\half y\right)\,
\left(\hat{x}^{\mu}\hat{t}^{\nu}+\hat{x}^{\nu}\hat{t}^{\mu}\right)
+(1-y)\,\hat{t}^\mu\hat{t}^\nu\biggr].
\label{zus}
\end{eqnarray}
The four tensor structures are mutually orthogonal.

The interesting physics is in the hadron tensor which, in leading order
in $1/Q$, is given by
\begin{equation}\label{jet}
2M {\cal W}^{\mu\nu}=2e^2\int {\md}^4k\,{\md}^4 k'\;\delta^4(k+q-k')\;\tr
\left[ \Phi(k)\, \gamma^\mu\, \Delta(k')\, \gamma^\nu \right].
\end{equation}
We consider for the moment only the quark contribution for one flavor.
Furthermore we limit ourselves to the symmetric part because the
unpolarized lepton tensor is symmetric. One has on
the distribution side (involving the target hadron)~\cite{sope}
\begin{equation}
\label{gijs}
\Phi_{\alpha \beta}(k)=\int\frac{{\md}^4 x}{(2\pi)^4}\,
{\e}^{{\mi} k\cdot x}\,\langle P S |\overline{\psi}_\beta(0){\cal G}(0,x)
\psi_\alpha (x)|P S\rangle ,
\end{equation}
while on the fragmentation side
(involving the produced hadron)~\cite{coll82}
\begin{equation}\label{bok}
\Delta_{\alpha \beta}(k')=\sum_X\frac{1}{2}\int
\frac{{\md}^4 x}{(2\pi)^4}\, {\e}^{{\mi} k'\cdot x}\,
\langle 0|{\cal G}(0,x)\psi_\alpha (x)|P_h,X\rangle
\langle P_h,X|\overline{\psi}_\beta (0)|0\rangle .
\end{equation}
In the first equation a color-summation is understood, in the second a
color-average.
A link operator ${\cal G}(0,x)$ = ${\cal P} \exp
\left[-{\mi} g\int_0^x {\md} s^\mu A_\mu(s)\right]$ is inserted to make
the definitions color gauge-invariant.
With an appropriate choice of the path structure in the link
operator used in the above definitions and an appropriate choice
of gauge, the above expression for the hadronic tensor
${\cal W}^{\mu \nu}$ corresponds
to the Born graph in a diagrammatic expansion~\cite{tang94a}.

In order to analyse the hadronic tensor it is convenient to work in a
frame in which the  hadrons $H$ and $h$ are collinear
(see also~\cite{meng92}). We can define the rank-two tensors
\begin{eqnarray}
&& g_T^{\mu \nu} \equiv g^{\mu \nu} - n_+^\mu n_-^\nu
- n_+^\nu n_-^\mu, \\
&& \epsilon_T^{\mu \nu} \equiv \epsilon^{\mu\nu}_{\ \ \ \rho \sigma}
n_+^\rho n_-^\sigma,
\end{eqnarray}
using the null vectors, defined implicitly by
\begin{eqnarray}
&&P = \frac{Q}{\xbj \sqrt{2}}\,n_+ +\frac{M^2\xbj}{Q\sqrt{2}}\, n_-,\\
&&P_h = \frac{z\,Q}{\sqrt{2}}\,n_- + \frac{M_h^2}{zQ\sqrt{2}}\,n_+,
\end{eqnarray}
satisfying $n_+ \cdot n_- =1$.
In analogy to the Drell-Yan
process~\cite{tang94a}, we define a `transverse' vector as
$a_T^\mu\equiv g_T^{\mu\nu}a_\nu$, having only transverse components
$\bat$ in the above-mentioned collinear frames.
Note that in general the photon momentum $q$ has non-zero transverse
components $\bqt$. Also, the hadron spin vector is decomposed according
to
\begin{equation}
S=\lambda\frac{Q}{\xbj M \sqrt{2}}\,n_+
-\lambda\frac{\xbj M}{Q \sqrt{2}}\,n_- + S_T,
\end{equation}
with $\lambda$ the helicity and $S_T$ the transverse spin, satisfying
$\lambda^2+\bSt^2=-S^2= 1$.

A `perpendicular' vector we define as $a_\perp^\mu\equiv
g_\perp^{\mu\nu}a_{T\nu}=a_T^\mu-(\bat\cdot\bqt/Q^2)q^\mu$.
These have only two non-zero components $\bap$ in the frames where
the hadron and the virtual photon are collinear. So in general the
outgoing hadron has non-zero $\bPhp$, specifically,
$\bPhp\approx-z\bqt$.
The use of `$\approx$' means `up to corrections of (relative) order
$1/Q^2$'.
The target hadron has neither $T$- nor $\perp$-components.
In short, a transverse tensor is perpendicular to both $P$ and $P_h$,
whereas a perpendicular tensor is perpendicular to both $P$ and $q$.
The two types are connected by boosts of order
$|\bPhp|/Q$, so that at leading order they may be freely
interchanged. Only at ${\cal O}(1/Q)$ the differences become
important~\cite{tang94b}.

After this kinematical intermezzo, we return to the calculation of the
hadronic tensor~(\ref{jet}). Assuming quark momenta
in hadrons to be limited, i.e., in the expressions for $\Phi$ and
$\Delta$ the quantities $k^2$, $k\cdot P$, $k'^2$, and $k'\cdot P_h$,
are of hadronic scale, one infers that $k^+\gg k'^+$ and
$k^-\ll k'^-$, so that
\begin{equation} \label{schapen}
\delta^4(k+q-k')\approx
\delta(k^++q^+)\,\delta(q^--k'^-)\,\delta^2(\bkt-\bkat-\bqt),
\end{equation}
such that $k^+/P^+ \approx -q^+/P^+ \approx \xbj$ and $P_h^-/k^{\prime
-} \approx  P_h^-/q^- \approx z$. Another consequence of
Eq.~(\ref{schapen}) is that one becomes sensitive to the integrals
$\int {\md} k^-\, \Phi(k)$ and $\int {\md} k^{\prime +}\,
\Delta(k^\prime)$.
Only specific Dirac projections will contribute in leading order.
For the distributions we need for polarized hadrons the matrix elements
($i=1,2$)
\begin{eqnarray}
\frac{1}{2}\int {\md} k^- \,\tr\left[\gamma^+\,\Phi(k)\right]&=&
f_1(x,\bkt^2),\label{eq:f1}\\
\frac{1}{2}\int {\md} k^- \,\tr\left[\gamma^+\gamma_5\,\Phi(k)\right]
&=&g_{1L}(x,\bkt^2)\,\lambda
+g_{1T}(x,\bkt^2)\frac{\bkt\cdot\bSt}{M},\label{eq:g1}\\
\frac{1}{2}\int {\md} k^-
\,\tr\left[{\mi}\sigma^{i+}\gamma_5\,\Phi(k)\right]&=&h_{1T}(x,\bkt^2)
\bSt^i +\left[h_{1L}^\perp(x,\bkt^2)\lambda
+h_{1T}^\perp(x,\bkt^2)\frac{\bkt\cdot\bSt}{M}\right]\frac{\bkt^i}{M},
\label{eq:h1}
\end{eqnarray}
where $x$ = $k^+/P^+ \approx \xbj$. Performing the $k_T$-integration,
only the distribution functions $f_1$, $g_{1L}$, and
$h_{1T}+(\bkt^2/2M^2)\,h_{1T}^\perp$,  contribute to the distributions
$f_1(x)$, $g_1(x)$, and $h_1(x)$, respectively.

Similarly, the leading parts for the (fragmentation) matrix elements in
Eq.~(\ref{bok}) for the case that one sums over the polarization of the
produced hadron are  ($i,j=1,2$)
\begin{eqnarray}
\frac{1}{2z}\int{\md} k^{\prime +}\,\tr\left[\gamma^-\,\Delta(k')\right]
&=& D_1(z,z^2\bbox{k}_T^{\prime 2}),\label{wim}\\
\frac{1}{2z}\int {\md} k^{\prime +}
\,\tr\left[{\mi}\sigma^{i-}\gamma_5\,\Delta(k')\right]
&=& \frac{\epsilon_T^{ij}\bbox{k}^\prime_{Tj}}{M_h} H_1^\perp(z,z^2
\bbox{k}_T^{\prime 2}),
\label{eq:H1}
\end{eqnarray}
where $z=P_h^-/k^-$ is the longitudinal momentum fraction of the
produced hadron, and $\bpt=-z\bkat$ is its transverse momentum with
respect to the fragmenting quark.
The normalization of $D_1$ is given by the momentum sum rule
$\sum_h \int_0^1 dz\int{\md}^2\bpt\,zD_1(z,\bpt^2$) = 1.
Hermiticity and parity invariance have been used in deriving the above
(real) parametrizations.
Interestingly, a structure corresponding to $H_1^\perp$ in
the distribution part will be excluded because of time-reversal
invariance.
The fragmentation functions are not invariant under the time-reversal
operation because the states $| P_h, X \rangle$ are out-states.
That $H_1^\perp$ can well be
nonzero in QCD has been made plausible in Refs.~\cite{coll93,coll94} by
applying simple  models. The function also appears in the cross section
for scattering of polarized leptons off unpolarized hadrons,
but in that case it is suppressed by a factor $1/Q$~\cite{leve94}.
Note that we use somewhat different functions as in Ref.~\cite{coll93},
the  connection being
$zD_1(z,z^2\bbox{k}_T^{\prime 2})=\hat{D}(z,k^\prime_T)$ and
$\epsilon_T^{ij}\bbox{k}^\prime_{Tj}\,zH_1^\perp(z,z^2
\bbox{k}_T^{\prime 2})/M_h=\Delta\hat{D} (z,k^\prime_T,e_T)$,
where $e_T$ is a unit vector in the $i$-direction.

By means of Fierz transformation of the two $\gamma$-matrices in
Eq.~(\ref{jet}), one finds that the projection $D_1$ in~(\ref{wim})
selects from the distribution side the function $f_1$ (called $\hat{f}$
in Ref.~\cite{coll93}). On the other hand, $H_1^\perp$ comes with the
matrix element in Eq.~\ref{eq:h1}. In the parametrization of this matrix
element three functions come in. In Ref.~\cite{coll93} only the function
$h_{1T}$ (called $\hat{f}_T$) was considered. We will consider here the
additional function $h_{1T}^\perp$ for transversely polarized hadrons
and $h_{1L}^\perp$ for longitudinally polarized hadrons.
The rest of the calculation is a matter of inserting the above
parametrizations into the (Fierz transformed) trace in Eq.~(\ref{jet}).
The result is
\begin{eqnarray}
2M\, {\cal W}^{\mu\nu} && =-2e^2\int {\md}^2\bkt
{\md}^2\bpt\,\delta^2\left(\bpt+z\bkt+z\bqt\right)
\nonumber\\
\times&&\biggl\{
f_1(x,\bkt^2)\,zD_1(z,\bpt^2)\,g_T^{\mu\nu}\nonumber\\
&&+\frac{h_{1T}(x,\bkt^2)\,H_1^\perp(z,\bpt^2)}{M_h}
\left[S_T^{\{\mu}\epsilon_T^{\nu\}\rho}p_{T\rho}
-(\epsilon_T^{\rho\sigma}S_{T\rho}p_{T\sigma})g_T^{\mu\nu}\right]
\nonumber\\
&&+\frac{\bkt\cdot \bSt}{M}\,
\frac{h_{1T}^\perp(x,\bkt^2)\,H_1^\perp(z,\bpt^2)}{M M_h}
\left[k_T^{\{\mu}\epsilon_T^{\nu\}\rho}p_{T\rho}
-(\epsilon_T^{\rho\sigma}k_{T\rho}p_{T\sigma})g_T^{\mu\nu}\right]
\nonumber\\
&& +\lambda\frac{h_{1L}^\perp(x,\bkt^2)\,
H_1^\perp(z,\bpt^2)}{M M_h}
\left[k_T^{\{\mu}\epsilon_T^{\nu\}\rho}p_{T\rho}
-(\epsilon_T^{\rho\sigma}k_{T\rho}p_{T\sigma})g_T^{\mu\nu}\right]
\biggr\}.
\end{eqnarray}
The way to project the Lorentz indices of the convolution variables
$k_T$ and $p_T$ on external momenta
was discussed in Appendix A of Ref.~\cite{tang94a} for the Drell-Yan
process. The resulting leading-order hadronic tensor is most
conveniently expressed in terms of four structure functions constructed
from the perpendicular vectors and tensors defined earlier,
\begin{eqnarray}
{\cal W}^{\mu\nu}=&&-{\cal W}_T\,g_\perp^{\mu\nu}-{\cal U}^T
\left(S_\perp^{\{\mu}\epsilon_\perp^{\nu\}\rho}\hat{h}_{\rho}
-(\epsilon_\perp^{\rho \sigma}S_{\perp \rho}\hat{h}_{\sigma})
g_\perp^{\mu\nu}\right)
\nonumber\\&& +({\cal V}^T\,\hh\cdot S_\perp-{\cal V}^L\,\lambda)
\left(\hh^{\{\mu}\epsilon_\perp^{\nu\}\rho}\hat{h}_{\rho}\right),
\label{eq:ht}
\end{eqnarray}
where $\hh=P_{h\perp}/|\bPhp|$.
Defining the convolution product (reinstating the sum over quark and
antiquark flavors)
\begin{eqnarray}
I[f D] & \equiv & \sum_a e_a^2\int {\md}^2\bkt
{\md}^2\bpt\,\delta^2\left(\bpt+z\bkt+z\bqt\right)\,f^a(\xbj,\bkt^2)
D^a(z,\bpt^2),\nonumber \\& \approx & \sum_a e_a^2\int {\md}^2\bkt
\,f^a(\xbj,\bkt^2)D^a\left(z,(\bPhp-z\bkt)^2\right),
\end{eqnarray}
where the use of transverse two-vectors is most appropriate because of
the definitions of the fragmentation functions. For the second line we
used $\bqt\approx -\bPhp/z$. The structure functions now read
\begin{eqnarray}
&&M{\cal W}_T=z\,I[f_1\,D_1],\label{sf}\\
&&M{\cal U}^T=I\Biggl[(\bh\cdot\bpt)\frac{h_{1T}\,H_1^\perp}{M_h}
\nonumber\\&&\qquad\qquad\quad
-\Bigl(2(\bkt\cdot\bh)^2(\bpt\cdot\bh)
-(\bkt\cdot\bh)(\bkt\cdot\bpt)
-(\bpt\cdot\bh)\, \bkt^2\Bigr)\frac{h_{1T}^\perp H_1^\perp}{M^2 M_h}
\Biggr],\label{UT}\\
&&M{\cal V}^T=I\Biggl[\Bigl(4(\bkt\cdot\bh)^2(\bpt\cdot\bh)
-2(\bkt\cdot\bh) (\bkt\cdot\bpt)
-(\bpt\cdot\bh)\, \bkt^2\Bigr)
\frac{h_{1T}^\perp\,H_1^\perp}{M^2 M_h}\Biggr],\label{VT}\\
&&M{\cal V}^L=I\Biggl[\left(2(\bkt\cdot\bh)(\bpt\cdot\bh)
-\bkt\cdot\bpt\right)\frac{h_{1L}^\perp\,H_1^\perp}{M M_h}\Biggr].
\label{VL}
\end{eqnarray}
The dot products can easily be converted into squares $\bkt^2$,
$\bpt^2$, and $\bPhp^2$, using the delta function
in the convolution integrals. So the structure functions depend on the
variables $\xbj$, $z$, and $\bPhp^2$.
Finally, the leading-order cross section is obtained by contracting
Eq.~(\ref{eq:ht}) with the leptonic tensor, Eq.~(\ref{zus}),
\begin{eqnarray}
\frac{{\md}\sigma}{{\md}\xbj\,{\md} y\,{\md} z\,{\md}^2 \bPhp}
&&=\frac{4\pi \alpha^2}{yzQ^2}\biggl[ {\cal W}_T (1-y+\half y^2)
+{\cal U}^T |\bSt|\,(1-y)\sin(\phi+\phi_S)\nonumber\\
&&\qquad\qquad+{\cal V}^T|\bSt|\,(1-y)\cos\phi_S\sin 2\phi
+{\cal V}^L\lambda\,(1-y)\sin 2\phi\biggr],
\end{eqnarray}
where $\cos\phi=-\hat{x}\cdot \hh$ and
$\sin\phi=\epsilon_{\perp}^{\mu\nu}\hat{x}_{\mu} \hat{h}_{\nu}$,
and likewise for the azimuth of the spin vector,
$|\bSp|\,\cos\phi_S=-\hat{x}\cdot S_\perp$ and
$|\bSp|\,\sin\phi_S=\epsilon_{\perp}^{\mu\nu}\hat{x}_{\mu}S_{\perp\nu}$
and we used $|\bSp|\approx |\bSt|$.
The expression for the structure function ${\cal U}^T$, Eq.~(\ref{UT}),
shows that the $\sin(\phi + \phi_S)$ asymmetry in the process
$e+p^{\uparrow}\rightarrow e+h+X$, found in Ref.~\cite{coll93}, not only
probes the transverse spin distribution $h_{1T}(x,\bkt^2)$ but also the
distribution $h_{1T}^\perp(x,\bkt^2)$. The latter function, however, is
independently probed in the $\sin 2\phi$ asymmetry for transversely
polarized hadrons [Eq.~(\ref{VT})]. The occurrence of the latter angular
dependence in the distribution of produced hadrons from a transversely
polarized target would therefore be a clear sign of the presence of this
new function.
In addition, another distribution function $h_{1L}^\perp(x,\bkt^2)$ is
probed in the $\sin 2\phi$ asymmetry for longitudinally polarized
hadrons [Eq.~(\ref{VL})].
These three distribution functions determine the transverse spin density
in a polarized hadron including the dependence on quark transverse
momenta as given in Eq.~(\ref{eq:h1}).
They are all convoluted with the same fragmentation function
$H_1^\perp(z,\bpt^2)$.

On completion of this work, we became aware of Ref.~\cite{kotz94}
treating semi-inclusive deep-inelastic scattering in a similar fashion.
Although the author also gets $\sin(\phi+\phi_S)$ and $\sin 2\phi$
single-spin asymmetries, he assumes exponential $\bkt$-behavior for
the distribution and fragmentation functions.
Our expressions for the structure functions, Eqs.~(\ref{sf})-(\ref{VL}),
are valid for {\em any\/} $\bkt$-dependence.

We acknowledge discussions with J.C. Collins.
This work is part of the research program of the foundation for
Fundamental Research of Matter (FOM) and the National Organization
for Scientific Research (NWO).

\end{document}